\documentclass{PoS}

\title{Exoplanets: Possible Biosignatures}

\ShortTitle{Possible Biosignatures}

\author{\speaker{R. Claudi}\\
        I.N.A.F. Osservatorio Astronomico di Padova\\
        E-mail: \email{riccardo.claudi@oapd.inaf.it}}

\abstract{The ancestor philosophers' dream of thousands of new worlds is finally realised: about 3500 extrasolar planets have been discovered in the neighborhood of our Sun. Most of them are very different from those we used to know in our Solar System. Others orbit their parent star inside the belt known as Habitable Zone (HZ) where a rocky planet with the appropriate climate could have the availability of liquid water on its surface. Those planets, in HZ or not, will be the object of observation that will be performed by new space- and ground-based instrumentation.
Space missions, such as JWST and the very recently proposed ARIEL (ESA M-Class Mission), or ground based instruments, like SPHERE@VLT, GPI@GEMINI and EPICS@ELT, have been proposed and built to measure the atmospheric transmission, reflection and emission spectra over a wide wavelength range.
Exoplanets are unique objects in astronomy because they have local counterparts the Solar System planets available for comparative planetology studies, but there are also really interesting outsider cases like super Earths. In our own system, proto-planet evolution was flanked by an active prebiotic chemistry that brought about the emergency of life on the Earth. The search for life signatures requires as the first step the knowledge of planet atmospheres, main objective of future exoplanetary space explorations. Indeed, the quest for the determination of their chemical composition is of much larger value than suggested by the specific case. It opens out to the more general speculation on what such detection might tell us about the presence of life on those planets. As, for now, we have only one example of life in the universe, we are bound to study terrestrial organisms to assess possibilities of life on other planets and guide our search for possible extinct or extant life on other planetary bodies.
The planet atmosphere characteristics and possible biosignatures will be inferred by studying such composite spectrum in order to identify the emission/absorption lines/bands from atmospheric molecules as water (H$_2$O), carbon monoxide (CO), methane (CH$_4$), ammonia (NH$_3$) etc.}

\FullConference{Frontier Research in Astrophysics – II\\
		23-28 May 2016\\
		Mondello (Palermo), Italy}

\begin{document}

\section{Introduction}
\label{sec:intro}
At the beginning of the second half of the previous century, Giuseppe Cocconi and Phillip Morrison published the first article (\cite{cocconiandmorrison1959}) devoted to a scientific way to search for alien life. They proposed to search for advanced civilizations wishing to communicate with other intelligent forms of life in the Galaxy searching for radio signals in the \emph{ neighborhood of 1,420 Mc/s}. They concluded their seminal article with the exhortation: \emph{the probability of success is difficult to estimate, but if we never search the chance of success is zero}.

We know the story. From here the SETI project was born (\cite{drake1961}, for a review see \cite{tarter2001}) together with its expectations and eeire silences (\cite{davies2010}).

In the modern view, the search for an intelligent and advanced civilization fades to leave space to search for signs of both existed and extant alien life tout court. In any case there are three main questions we have to answer to begin the hunt: where? what? how?

To answer the first question, the bodies of Solar System are  the first place where to search. Venus and Mars are the easiest targets for this quest, but also the icy moons of the giant planets, Europa, Enceladus and Titan could be rummaged to search for life. In addition, since 1995, year of discovery of the hot Jupiter 51 Peg b (\cite{mayorandqueloz1995}), a multitude of new worlds where to search for signs of life have been discovered (about 3500 planets \cite{schneideretal2011}). 

The second question is the hardest. What kind of life we are looking for? The simplest answer is: we search for life as we know it.   Life has been described as a (thermodynamically) open system (\cite{prigogineetal1972}), which exploits gradients in its surroundings to create imperfect copies of itself, makes use of chemistry based on carbon, and exploits liquid water as solvent for the necessary chemical reactions (\cite{owen1980}, \cite{desmaraisetal2002}). The search for this specific form of life seems a priori statements, but considering life like a phenomenon with a non--zero probability to happen as soon as the conditions are satisfied it appears, we have to consider all the conditions that maximize this probability. In this framework, carbon is the only atom with which it is possible to form molecules with a total  number of atoms up to 13 (e.g. HC$_{11}$N) thus allowing the formation of very complex molecules. Carbon is also very easy to oxidize (CO$_2$) and reduce (CH$_2$).
For what concerns liquid water, it has some important characteristics that make it the best solvent for life: a large dipole moment, the capability to form hydrogen bonds, to stabilize macromolecules, to orient hydro--phobic--hydrophilic molecules, etc. Water is an abundant compound in our galaxy, it can be found in different environments, from cold dense molecular clouds to hot stellar atmospheres (e.g., \cite{cernicharoandcrovisier2005}, \cite{lammeretal2009}). Water is liquid at a large range of temperatures and pressures and it is a strong polar--nonpolar solvent. This dichotomy is essential for maintaining stable biomolecular and cellular structures (\cite{desmaraisetal2002}). Furthermore, liquid water has a great heat capacity that makes it able to tolerate a heat shock. In addiction, the most common solid form of water has a specific weight lighter than that of its liquid form. This allows ice to float on a liquid ocean safeguarding the underlying liquid water. All those characteristics let grow the probability that life, once it emerges, could survive and evolve.

So, our search for signs of life is based on the assumption that alien life shares fundamental characteristics with life on Earth. Life based on a different chemistry is not considered here because such life--forms, should they exist, would have by--products that are so far unknown.  

Over one-half a century ago, the approach to remote detection of signs of life on another planet was set out in \cite{lederberg1965} and \cite{lovelock1965}, which introduced the canonical concept for the search for an atmosphere with gases severely out of thermochemical redox equilibrium\footnote{Redox chemistry adds or removes electrons from an atom or molecule (reduction or oxidation, respectively). Redox chemistry is used by all life on Earth and thought to enable more flexibility than non--redox chemistry. }. The idea that gas by--products from metabolic redox reactions can accumulate in the atmosphere was initially favoured for future sign of life  identification, because abiotic processes were thought to be less likely to create a redox disequilibrium. We call biosignatures gases all those gases that are produced by life and that can accumulate in a planet atmosphere to detectable levels.

\begin{figure}[t]
\begin{center}
 \includegraphics[width=.8\textwidth]{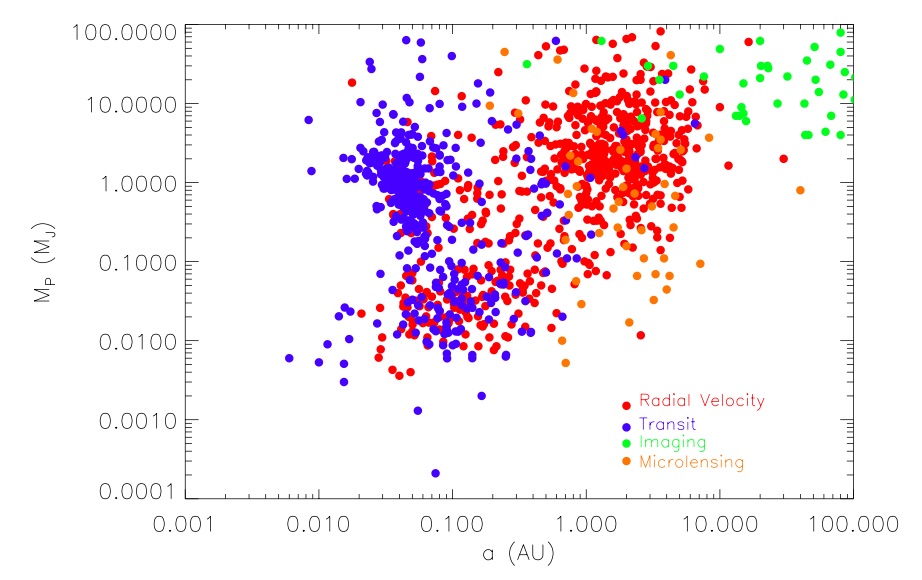}
\caption{Mass distribution as function of orbital distance of all extrasolar planets discovered up to 2016. The planets are distinguished by the discovering methods.}
 \label{fig:distramass}
\end{center}
\end{figure}

These considerations bring us in a natural way to answer to the third question. In the case of the bodies of the Solar System, the best way to investigate if they are bearing life on their surface is to go on--site and explore the planet and or its moons searching for evidences of life. This is made ''easy'' by the vicinity of these planets. Venus and Mars are targets of this wandering of the human beings in the Solar System. This is not possible yet for extrasolar planets. In this case, and also in the former actually, the remote sensing is the ''easier'' way of investigation. In the last years several different atmosphere observation techniques have been developed. These techniques have been used mainly to study the transiting Hot Jupiter atmospheres (for a review see e.g.\cite{seageranddeming2010}) and are going to be fine tuned in order to observe habitable super Earths using new instrumentation and space missions.

This paper will review the sites where to search for life and their main characteristics, what we should search for in their atmospheres (biosignatures) and why, and eventually the main methods for this quest.  

\section{Super Earths and Habitability}
The number of new worlds is continuously growing (see e.g. www.exoplanet.eu \cite{schneideretal2011}). So far, about 3500 confirmed planets and over than 2500 planet candidates have been found mostly with radial velocity and transit methods.  The distribution of mass of these new worlds includes planets more massive than Jupiter (super Jupiter) but also planets that fill the mass gap between Mars and Neptune masses (Earths and super Earths, mini Neptunes) unveiling populations of new kinds of planets completely unknown in the Solar System (see Fig\ \ref{fig:distramass}). 

Another striking feature of the distribution of observed planets is that a lot of them orbits their star at a distance that is smaller than Mercury's distance from the Sun ($\sim 0.5$ au). These planets are strongly irradiated by their host star and assume the adjective of ''hot''.

\begin{figure}
\begin{center}
 \includegraphics[width=.7\textwidth]{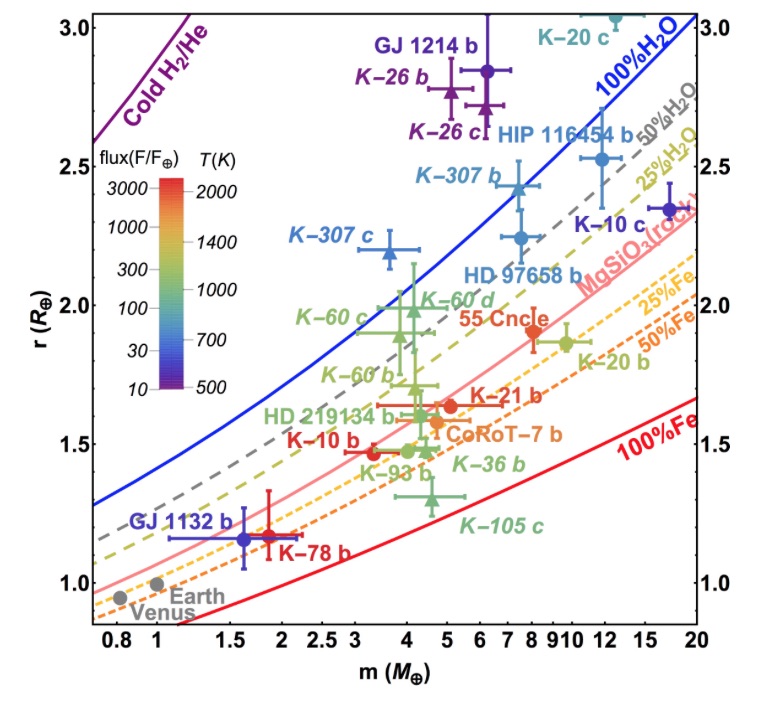}
\caption{The Mass Radius plane in the super Earth and mini neptune region (up to 20 M$_\oplus$). The continuous lines correspond to density of planets with different compositions. Circles are for masses measured by radial velocity method, triangles are for masses measured by TTV (transit time variations). The color code defines the bolometric stellar flux (in Earth fluxes) impinging on the planets (taken by \cite{zengetal2016} or www.cfa.harvard.edu/$\sim$lzeng/)}
 \label{fig:massaR}
\end{center}
\end{figure}

\begin{figure}[h]
\begin{center}
 \includegraphics[width=.8\textwidth]{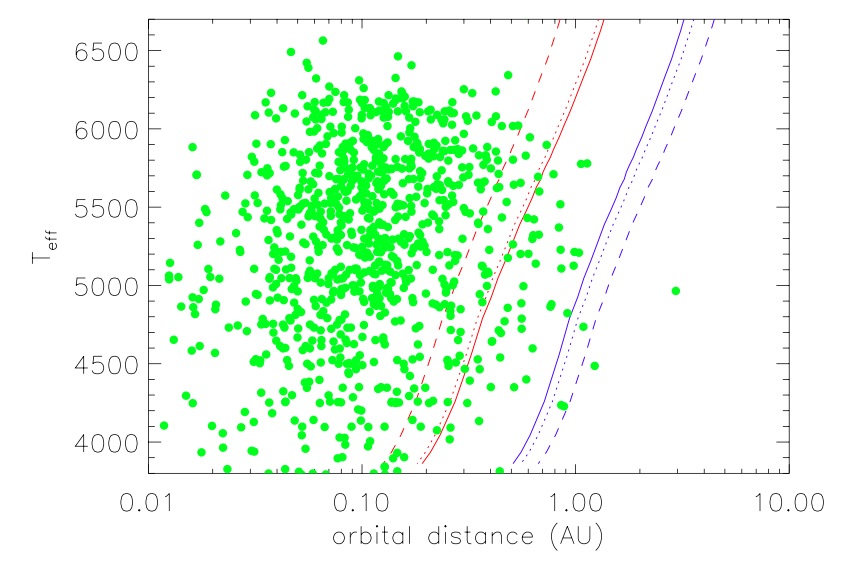}
\caption{Orbital distance $a$ of extrasolar planets with masses ranging in the $1\leq \ M_\oplus \ \leq10$ and the T$_{eff}$ of their host stars compared with the HZ as defined by Selsis et al. \cite{selsisetal2007} for stars with different spectral type. The inner (red lines) and outer (blue lines) limits of HZ corresponds to ''recent Venus'' and  ''early Mars'' criteria (continuos line), and the radiative-convective models with a cloudiness of 50 (dotted line), and 100\% (dashed line).  The data on super Earths were taken and merged between the exoplanets encyclopaedia and NASA extrasolar planets database.}
 \label{fig:searth_hz}
\end{center}
\end{figure}

\noindent
The largest population of exoplanets is constituted by one kind of these never seen planets: super Earths.  These are rocky exoplanets with mass ranging between 1 and 10 M$_\oplus$ (\cite{valenciaetal2007}). Although the lower mass limit is obvious for historical reasons, the upper limit is somewhat arbitrary. It is due to the physical argument that above 10 M$_\oplus$, planets can retain hydrogen and helium in their atmospheres (\cite{idaelin2004}). This class of small mass planets is also characterised by radii ranging between 1 and 2 R$_\oplus$ (\cite{batalhaetal2013}; \cite{boruckietal2011} and \cite{fressinetal2013}).
The first super--Earth with a mass lower than 10 M$_\oplus$ was discovered by \cite{riveraetal2005}, orbiting an M4 V star named GJ 876. Its estimated mass is $(7.5 \pm 0.7)$ M$_\oplus$ and it has an orbital period of 1.94 days. It is close to the host star, and the surface temperature is calculated to lie between 430 and 650 K (\cite{riveraetal2005}). The improvements in radial velocity precision together with well focused observational strategies brought the discovery of a large number of low mass planets around M stars with an occurrence of 40\% for stars that host planets with minimum mass between 3 and 30 M$_\oplus$ and P$<50d$ (\cite{mayoretal2014}).  This population doesn't emerge only from the radial velocity method score but also in the midst of the \emph{Kepler} results.  A lot of small planets have been found transiting their host star with radii between those of Earth and Neptune ($1-3.8$ R$_\oplus$) and they show also a large range of densities (see Figure\ \ref{fig:massaR}, \cite{zengetal2016}, \cite{lissaueretal2014}).  

These high occurrence rates have one consequence: systems of multiple planets with masses between 1 M$_\oplus$ and 20 M$_\oplus$ orbiting within 0.5--1.0 au represent the most common type of planetary systems in the Galaxy (\cite{mayoretal2014}). 
The importance of super--Earth, either they are similar to Earth or not, is that they are rocky planets with a rigid interface between the interior and the atmosphere and if they stay at the right distance from their host star, they could retain liquid water on their surface. This condition generally defines a planet as habitable (because all life on Earth requires liquid water). Surface liquid water requires a suitable surface temperature. The surface temperature of planets with thin atmospheres is determined by the fraction of flux reaching the surface of the planet from the host star. Planets at the ''right distance'' are planets on orbits between the hot planet and the cold planet orbits, in particular they are in the well known ''Habitable Zone'' (HZ \cite{kastingetal1993}; \cite{selsisetal2007}, \cite{koppaparuetal2013}), or the zone around a star where a rocky planet with a thin atmosphere, heated by its star, may have liquid water on its surface (see Figure\ \ref{fig:searth_hz}). The HZ position depends on the star: small stars with lower luminosity have HZ closer to them compared to larger stars. Also the climate of the planet has an impact on the surface temperature, in fact it is the greenhouse warming of rocky planet atmospheres that controls the surface temperature. The revised view is that planet habitability is planet-specific, because the huge range of planet diversity in terms of masses, orbits, and star types should extend to planet atmospheres based on the stochastic nature of planet formation and subsequent evolution. 
On the other hand the condition that makes a planet habitable is much more complex than having a planet located at the right distance from its host star. Furthermore, various geophysical and geodynamical aspects, the radiation and the host stars plasma environment can influence the evolution of Earth and super--Earths and life if it originated (\cite{scaloetal2007}; \cite{lammeretal2007}). 
The HZ is not the only site where is possible to have liquid water. For instance, a subsurface ocean within the satellite of a gas giant may be habitable for some alien life form, not necessarily for life as we know it. On the other hand there may be many habitable exoplanets, on which life could have  existed or is still extant, or planets that could have been habitable once but do not bear any life at all. The HZ concept is an useful one, especially for remote sensing, but it is more important to keep an open mind and search for every weird thing and out of equilibrium quantities and eventually investigate them.

\section{Biosignatures}

The term ''Biosignature'' means detectable atmospheric gas species, or a set of species, whose presence at significant abundance strongly suggests a biological origin (\cite{desmaraisetal2002}). Other gases that are or could be indicative of biological processes but can also be produced abiotically, are called bioindicators (e.g. on Earth, O$_3$ is photochemically produced by O$_2$). Their quantity and detection, along with other atmospheric species, all within a certain context (for instance, the properties of the star and the planet) points toward a biological origin. These gases should be ubiquitous by--products of carbon--based biochemistry, even if the details of alien biochemistry are significantly different from the biochemistry on Earth. Really complete reviews on biosignature types are given in \cite{desmaraisetal2002}, \cite{seagerandbains2015}, \cite{kalteneggeretal2010a}, \cite{seguraandkaltenegger2008},  \cite{seageretal2012}, \cite{seageretal2013} where the reader can find precise information on this topic. In the following the main results of these authors will be summarized.

The starting point in understanding how to search for life on exoplanets with remote sensing (i.e studying their atmospheres) begins with Earth, the only planet, as we know so far that hosts life. So the search for signs of life is based on the assumption that extraterrestrial life has the fundamental characteristics of life as we know it or, in other words, it mainly requires liquid water as a solvent and has a carbon-based chemistry (see, e.g.,\cite{desmaraisetal2002},  \cite{brack1993}). 

During the second half of the previous century a lot of studies put the focus on this topic. The very first discussions on the search for life by atmospheric sensing focused on the idea that a system at thermodynamic disequilibrium shall be by itself an atmospheric biosignature (\cite{lederberg1965}, \cite{lovelock1965}). This disequilibrium (sign of life) could be testified by the simultaneous presence in the Earth's atmosphere of hydrocarbons and molecular oxygen (a redox disequilibrium). Lippincot and collaborators (\cite{lippincotetal1967}) identified Methane (CH$_4$) as the hydrocarbon to be contrasted with O$_2$ as sign of life in their first systematic thermodynamic equilibrium calculations on the atmospheres of the rocky planets of the Solar System. They found that CH$_4$ is out of thermodynamic equilibrium on Earth together with other gases (H$_2$, N$_2$O, and SO$_2$) but they all cannot be considered unambiguous signs of life (with the possible exception of N$_2$O) due to the production of these molecules by geochemical processes. In the 1975, Lovelock et al (\cite{lovelocketal1975}) supported   the idea of Lippincot et al. \cite{lippincotetal1967} that the O$_2$--CH$_4$ disequilibrium was strong evidence for life  after that CH$_4$ as a biosignature gas also seemed to be established (\cite{lovelocketal1975}, \cite{sagan1975}). In any case, a lot of arguments tackles the use of thermodynamic equilibrium as a life indicator (\cite{seager2014}). The first point is that almost any gas other than N$_2$ and CO$_2$ in Earth's atmosphere, however generated, is out of thermodynamic equilibrium because of the Earth's high O$_2$ levels. So, the argument that Earth's atmosphere is out of thermodynamic equilibrium reduces to a statement about the high levels of Earth's atmospheric O$_2$. Also if no or too little O$_2$ is present it is possible to have significant thermodynamic disequilibrium due to geochemical or photochemical processes.

\begin{figure}
\begin{center}
 \includegraphics[width=.7\textwidth]{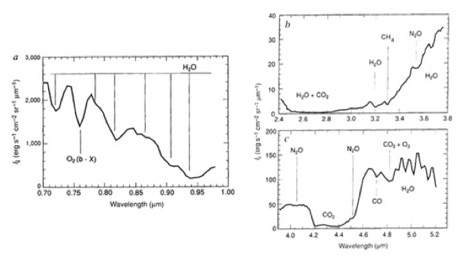}
\caption{The original Earth spectrum taken by Galileo fly--by on Earth (\cite{saganetal1993}).}
 \label{fig:galearthspec}
\end{center}
\end{figure}

During its mission, flying towards Jupiter, the probe Galileo took UV, visible and NIR spectra of Earth that Sagan et al. (\cite{saganetal1993}) analyzed searching for signatures of life.  They found a large amount of O$_2$ and the simultaneous presence of CH$_4$ traces concluding that this co--presence is strongly suggestive of life (see Fig\ \ref{fig:galearthspec}). We can consider this as the first remote sensing observation of a planet in the quest for life. 

So, considering Earth as an exoplanet, it is possible to observe spectra with a relatively prominent oxygen absorption feature at 0.76 $\mu$m, whereas methane (at present day levels of 1.6 ppm) has only extremely weak spectral features. Moreover, on Earth, some atmospheric species that exhibit observable spectral features come directly or indirectly from biological activity. The main molecules are O$_2$, O$_3$, CH$_4$, and N$_2$O. Furthermore both CO$_2$ and H$_2$O are important greenhouse gases and also potential sources for high O concentration from photosynthesis.

The chemicals produced by life on Earth are hundreds of thousands (estimated from plant natural products, microbial natural products, and marine natural products), but only a subset of hundreds are volatile enough to enter the atmosphere at more than trace concentrations. Among these, only a few handfuls accumulate to high enough levels to be remotely detectable for astronomical purposes and defined as biosignatures. 
Apart from oxygen, these biosignature (and bioindicator) gases range from highly abundant gases in Earth's atmosphere that are either already existing or predominantly produced by geochemical or photochemical processes (N$_2$, Ar, CO$_2$, and H$_2$O) to those that are relatively abundant and attributed to life (N$_2$O, CH$_4$ and H$_2$S). We have to consider also gases that are weakly present but may play important roles in the atmospheric processes (DMDS and CH$_3$Cl) and gases that are present only in trace amounts including the hundreds of minutely present volatile organic compounds released by trees in a forest or fungi in the soil (\cite{seagerandbains2015}). 

\begin{table}[h]
\begin{center}
\begin{tabular}{cccl}
\hline
Reductant         & Oxidant          & Output                & Comment \\
\hline
\hline
\multicolumn{3}{l}{\it Oxidation of Organic Matter}        &         \\
CH$_2$O         &  O$_2$          & CO$_2$, H$_2$O    &    \\
\hline
\multicolumn{3}{l}{\it Hydrogen Oxidation}                     &         \\
H$_2$              &  O$_2$          &  H$_2$O                   &          \\
H$_2$              &  H$_2$O$_2$          &  H$_2$O        &          \\
\hline
\multicolumn{3}{l}{\it Sulfur Compound Oxidation}        &         \\
H$_2$S            &  O$_2$          &  SO$^{2-}_4$              &          \\
HS$^-$             &  O$_2$          &  S                               &          \\
S                      &  O$_2$           &  SO$^{2-}_4$              &          \\
S$_2O^{2-}_4$ &  O$_2$          &  SO$^{2-}_4$              &          \\
\hline
\multicolumn{3}{l}{\it Iron Oxidation }        &         \\
Fe$^{2+}$         &   O$_2$          &  Fe$^{3+}$, OH$^-$   &       \\
\hline
\multicolumn{3}{l}{\it Ammonia Oxidation }        &         \\
NH$_3$                &  O$_2$           &  NO$^-_2$, H$_2$O   &  Acqueous (Nitrite)     \\
NH$^+_4$            &  O$_2$          &  NO$^-_2$, H$_2$O &  Acqueous (Nitrite)       \\
NO$^-_2$            &  O$_2$          &  NO$^-_3$                 &  Biological/Abiological     \\
\hline
\end{tabular}
\caption{Aerobic Chemotrophy: summary of redox reaction by-products}
\label{tab:biouno}
\end{center}
\end{table}%

To make some order in this matter, it is useful to divide the different possible biosignatures considering how they are produced by life. The main processess that we have to take into account are: the gathering of energy by living beings in order to use it to live (metabolic reactions), the construction of organic matter and the production of gases for reasons that are specific of the organism. Briefly: biosignature of type I, type II and type III (\cite{seageretal2013}).

\begin{table}[h]
\begin{center}
\begin{tabular}{cccl}
\hline
Reductant         & Oxidant          & Output                & Comment \\
\hline
\hline
\multicolumn{3}{l}{\it Denitrification }        &         \\
H$_2$                  &NO$^-_3$      &NO$^-_3$, H$_2$O &   Biological/Abiological   \\
H$_2$                  &NO$^-_2$      &NO, H$_2$O & Weak spectral feature      \\
H$_2$                  &NO                 &N$_2$O, H$_2$O & Weak spectral feature      \\
H$_2$                  &N$_2$O         &N$_2$, H$_2$O & Metabolic product      \\
Fe$^{2+}$            &NO$^-_3$      &NO$_2$, Fe$^{3+}$  & Weak spectral feature    \\
Fe$^{2+}$            &NO$^-_2$      &NO, Fe$^{3+}$  & Weak spectral feature    \\
Fe$^{2+}$            &NO                 &N$_2$O, Fe$^{3+}$  & Weak spectral feature    \\
Fe$^{2+}$            &N$_2$O         &N$_2$, Fe$^{3+}$  & Metabolic product     \\
\hline
\multicolumn{3}{l}{\it Iron Reduction }        &         \\
Organics             & Fe$^{3+}$  & Fe$^{2+}$  & Anaerobic bacteria, precipitating minerals  \\
H$_2$                 & Fe$^{3+}$  & Fe$^{2+}$, Fe$^{3+}$   & Precipitating minerals      \\
\hline
\multicolumn{3}{l}{\it Sulfur Reduction }        &         \\
Organics             & SO$^{2-}_4$              & SO$^{2-}_3$, SO$_2$, H$^+$, CO$_2$ &    \\
H$_2$                 & SO$^{2-}_4$              & SO$^{2-}_3$, SO$_2$, H$^+$ &      \\
H$_2$                 & SO$^{2-}_3$              & S$_2$O$_3$, H$^+$              &         \\
H$_2$                 &SO$^{-}_3$                 & S$^0$, H$^+$                          &         \\
H$_2$                 &SO$^0$                       & H$_2$S, H$^+$                       &          \\
CH$_4$               & SO$^{2-}_4$              & H$_2$S, CO$_2$                    &         \\
\hline
\multicolumn{3}{l}{\it Methanogenesis }        &         \\
Organics             & CO$_2$                        & CH$_4$, H$_2$O            &        \\
H$_2$                 & CO$_2$                        & CH$_4$, H$_2$O            & Abiotic Pathway       \\
\hline
\multicolumn{3}{l}{\it Anammox }        &         \\
NH$_3$              & NO$^-_2$                     & N$_2$, H$_2$O               &         \\
NH$^+_4$          & NO$^-_2$                     &  N$_2$, H$_2$O               &         \\
NH$_3$              & NO$^-_3$                     &  N$_2$, H$_2$O               &         \\
NH$^+_4$          & NO$^-_3$                     &  N$_2$, H$_2$O               &         \\
\hline
\hline
\end{tabular}
\caption{Anaerobic chemotrophy: summary of redox reaction by-products}
\label{tab:biosignatures2}
\end{center}
\end{table}%

\subsection{Primary metabolic biosignatures (Type I)}

The type I biosignature category contents those by--product gases produced from metabolic reactions that capture energy from environmental redox chemical potential energy gradients (\cite{seageretal2012}; \cite{seageretal2013}). Such gases (see Table\ \ref{tab:biouno} for aerobic chemotrophy and Table\ \ref{tab:biosignatures2} for anaerobic chemotrophy) are abundant due to the presence of large quantities of reactant in the environment, but they could be easily masked by false positives.  Geology, for example, works on the same molecules as life does. Moreover, in one environment, a given redox reaction will be kinetically inhibited and it is only started by life's enzymes, while in another environment with the right conditions (temperature, pressure, concentration and acidity), the same reaction might proceed spontaneously. 

A typical reaction of this category is the production of methane. It is considered a ''chemical energy gradient'' biosignature gas because it is generated by methanogen bacteria at the sea floor reducing the CO$_2$ available in the sea water due to the mixing with the atmosphere. In order to reduce the CO$_2$  these bacteria use the H$_2$ released by hot water coming from rocks (serpentinization). The reaction is:

$$H_2 + CO_2 \rightarrow CH_4 + H_2O $$

On the other hand methane is also released volcanically from hydrothermal systems. Most of the methane found in the present atmosphere of Earth has a biological origin but a small fraction is produced abiotically in hydrothermal systems where hydrogen is released by the oxidation of Fe by H$_2$O and reacts with CO$_2$ to form CH$_4$. The amount of CH$_4$ produced with this abiotic mechanism depends by the oxidation degree of the planetary crust. Therefore, the detection of CH$_4$ alone cannot be considered as a sign of life, though its detection in an oxygen--rich atmosphere could be an indication of the presence of a biosphere.
On early Earth, widespread methanogen bacteria (\cite{haqq--misraetal2008}) may have produced  CH$_4$  at much higher levels (1,000 ppm or even 1\%). Such high CH$_4$ concentrations would be easier to detect. On the other hand, at that time the contents of oxygen in the Earth's atmosphere was almost will, the O$_2$--CH$_4$ redox pairs would be challenging to detect concurrently (\cite{desmaraisetal2002}  and \cite{pavlovetal2003}), unless perhaps in the case of a planet in a lower-UV radiation environment (possible with some M host stars  \cite{seguraetal2005}). 
The recent confirmation of methane in the atmosphere of Mars (\cite{mummaetal2009}; \cite{websteretal2015}), that contains  0.1\% of O$_2$ and some O$_3$, is a good example for both the consideration of CH$_4$ as a biosignature gas, since it is photochemically unstable and must be actively produced, but it is also an example for a false positive because CH$_4$ could be produced geologically.

NH$_3$ is a very similar case to the one of CH$_4$. NH$_3$ is produced on Earth quite only by biological processes, apart from the one industrially manufactured.
They are both released into Earth's atmosphere by the biosphere with similar rates, but the atmospheric level of NH$_3$ is orders of magnitude lower due to its very short lifetime under UV irradiation. The detection of NH$_3$ in the atmosphere of a habitable planet would thus be extremely interesting, especially if found with oxidized species (\cite{kalteneggeretal2010b}) .

An interesting biosignature of this category is Nitrous oxide (N$_2$O). It is produced in abundance by life (denitrifying bacteria) but only in negligible amounts by abiotic processes. Most of the reaction that produce N$_2$O are listed in Table\ \ref{tab:biosignatures2}. It is difficult to be detected in a humid atmosphere because of the presence of the molecular band of water vapor while it would become more apparent in atmospheres with more N$_2$O or less H$_2$O vapor, or a combination of the two (\cite{kalteneggeretal2010b}). Segura et al.\ \cite{seguraetal2003} calculated the level of N$_2$O for different O$_2$ levels and found that, though N$_2$O is a reduced species compared to N$_2$, its level decreases with O$_2$. This is due to the fact that a decrease in O$_2$ produces an increase in H$_2$O photolysis, which results in the production of more hydroxyl radicals (OH) responsible for the destruction of N$_2$O.

\subsection{By--product of organic matter building (type II)}

The second category, type II biosignatures \cite{seageretal2013}, records few biosignature gases (see Table\ \ref{tab:phot}) among its elements. These gases are by--product of biomass building. These are mainly energy--requiring reactions that capture environmental carbon (and to a lesser extent other elements). On Earth the dominant example is O$_2$ produced by oxygenic photosynthesis, which gains energy from sunlight. In more detail, photosynthesis captures the carbon in CO$_2$ into biomass, releasing oxygen that, nowadays, is the 20\% by volume of the Earth's atmosphere. Less than 1ppm of atmospheric O$_2$ comes from abiotic processes (\cite{walker1977}). This high quantity of such a reactive gas like O$_2$ with a short atmospheric lifetime allows to consider oxygen a robust biosignature (\cite{legeretal1993}).
Owen (\cite{owen1980}) suggested searching for O$_2$ as a tracer of life. In fact, without continual replenishment by photosynthesis in plants and bacteria, O$_2$ would be ten orders of magnitude less than present today in the Earth's atmosphere (\cite{kastingandcatling2003}). Any observer seeing oxygen in Earth's spectrum would know that some non-geological chemistry must be producing it.

Photosynthesis efficiently converts light energy to electrochemical energy by redox reactions. Light, exciting pigments, causes a transfer of electrons along bio--chemical pathways having  the CO$_2$ reduction as result. The electron is replaced by one extracted from the reductant. The basic stoichiometry of photosynthesis is (\cite{kiangetal2007}):

$$ CO_2 + 2H_2X + h\nu \rightarrow (CH_2O) + H_2O + 2X $$

\noindent
This is a general way to show the reaction for both oxygenic and anoxygenic photosynthesis. H$_2$X represents the reductant that could be H$_2$O (oxygenic photosynthesis) or H$_2$S (an-oxygenic photosynthesis), h$\nu$ is the photon energy (h is the Planck's constant). In the case of oxygenic photosynthesis the reductant is water and we have the following reactions:

$$ 2H_2O + h\nu \rightarrow 4H^+ + 4e^- + O_2 $$  
$$ CO_2 + 4e^- + 4H^+ \rightarrow CH_2O + H_2O $$ 

Here the complete reaction is split in two parts: the former is the reaction where the electrons are generated, while the latter is  the synthesis of carbohydrate. The two processes, light capture and biomass building, are mechanistically distinct. For each O$_2$ molecule four photons are required (one photon for each bond in two water molecules) while other four photons are necessary to reduce CO$_2$. Thus, a minimum of eight photons is required both to evolve one O$_2$ and to fix carbon from one CO$_2$.

Cyanobacteria and plants are responsible for the production of oxygen by using solar photons to extract hydrogen from water (that is abundant on Earth) and using it to produce organic molecules from CO$_2$. The reverse reaction, using O$_2$ to oxidize the organics produced by photosynthesis, can occur abiotically when they are exposed to free oxygen or biotically by eukaryotes breathing O$_2$ and consuming organics. Because of this balance, the net release of O$_2$ in the atmosphere is due to the burial of organics in sediments. Each reduced carbon buried results in a free O$_2$ molecule in the atmosphere (\cite{seagerandbains2015}). This net release rate is also balanced by weathering of fossilised carbon when exposed to the surface. The oxidation of reduced volcanic gases, such as H$_2$ and H$_2$S, is also responsible for a significant fraction of the oxygen losses. The atmospheric oxygen is recycled through respiration and photosynthesis in less than 10,000 years (\cite{kalteneggeretal2010b}). In the case of a total extinction of Earth's biosphere, the atmospheric O$_2$ would disappear in a few million years.

The oxygenic phothosynthesis on Earth is a successful method to transform radiative energy in chemical energy an store it in organic matter. This happens on a planetary scale impacting on the host environment leading to a global transformation.

Anoxygenic photosynthesis uses instead other reductants, like for example H$_2$S, H$_2$ and Fe$^{2+}$. 

When the reductant is H$_2$S, elemental sulfur is produced instead of oxygen  (\cite{kiangetal2007}):

$$CO_2 + 2H_2S + h\nu \rightarrow (CH_2O) + H_2O + 2S$$
$$3CO_2 +2S + 5H_2O + h\nu \rightarrow 3(CH_2O) + 2H_2SO_4$$

where H$_2$S is split by photons to yield an electron donor, CH$_2$O represents the carbohydrates incorporated into the microbe, and S and H$_2$O are the metabolic by--products. Eventually, sulfur may be oxidized to sulfate which is not a gas and cannot enter the atmosphere as a biosignature. Also in this case the quantum requirement is  8 to 12 photons per carbon fixed. In summary, the inputs to photosynthesis are light energy, a carbon source and a reductant (see Table\ \ref{tab:phot}). The outputs are carbohydrates, elemental sulfur, water and other oxidised forms of the reductant in the reaction.

\begin{table}[h]
\begin{center}
\begin{tabular}{cccl}
\hline
Input        & Radiation         & Output                & Comment \\
\hline
\hline
\multicolumn{3}{l}{\it Oxigenic Photosynthesis}  &         \\
H$_2$O    & $h\nu$            &  O$_2$               &  Solid biosignature  \\
\hline
\multicolumn{3}{l}{\it Anoxygenic Photosynthesis}  &         \\
H$_2$S           &  $h\nu$            & S                     &          \\ 
S$_2$O$^-_3$  &  $h\nu$            &  H$_2$SO$_4$ &         \\
S                      & $h\nu$            &  H$_2$SO$_4$ &         \\
H$_2$              & $h\nu$            &  H$_2$O           &         \\
Fe$^{2+}$        & $h\nu$            & Fe$^{3+}$         &         \\
NO$^-_2$         & $h\nu$            & NO$^-_3$          &         \\
\hline
\hline
\end{tabular}
\caption{Phototrophy: summary of photosynthetic reaction by--products.}
\label{tab:phot}
\end{center}
\end{table}%

The best habitats for anoxygenic photosynthetic organisms are illuminated environments but with no free oxygen. Actually, for these organisms oxygen is a poison. Anoxygenic bacteria can be found in freshwater lakes and ponds, hot and sulfur springs, and some marine waters where the sources of electron donors (e.g., H$_2$S) can be either geological (in sulfur springs) or biological (produced by sulfate-reducing bacteria). From an evolutionary point of view, anoxygenic photosynthesis is believed to have preceded oxygenic photosynthesis and to have appeared on Earth more than 3 billion years ago (\cite{seageretal2012}, \cite{desmarais2000}).

\subsection{Secondary metabolic biosignatures (type III)}

A lot of chemical substances are synthetized by living organisms in order to answer to very different stimuli of the environment. These substances are highly specialized chemicals not directly tied to the local chemical environment and thermodynamics that are produced for reasons other than energy capture or the construction of the basic components of life. Among the main functions there are also the defense against the environment or against other organisms, signaling, or internal physiological control. For these reasons, type III biosignatures have much more chemical variety if compared with the other type of biosignatures (\cite{seageretal2012}, \cite{seageretal2013}). Some of molecules produced as by--products to primary metabolism are also produced by secondary metabolism. These include CH$_4$, NO, H$_2$S, and CO and are produced through different chemical routes with respect to those used in primary metabolism. On the other hand there are compounds and molecules that cannot enumerate in the other category of biosignatures. Most of them are inorganic compounds like sulfur and nitrogen compounds or organic molecules like isoprene and terpenoids (VOC or volatile organic carbon) and halogenated organics and all are potentially considered biosignatures (for a complete description see \cite{seageretal2012}). Also if they are generally expected to be produced in small quantities, the wider variety with respect to the other two types of biosignatures should make them to be less prone to false positives. Furthermore, because they require energy and specific catalysis to be produced, type III biosignature gases are unlikely to be geologically made in substantial amounts, so they are unlikely to be present in the absence of life.

On Earth some sulfur compounds released in atmosphere are very promising  in unveiling life as biosignature. The gases to be considered are hydrogen sulfide (H$_2$S, that is also produced by primary methabolism), carbon disulfide (CS$_2$), carbonyl sulfide (OCS, sometimes written as COS), dimethyl sulfide (DMS), and dimethyl sulfoxide (DMSO: CH$_3 \cdot$SO$_2 \cdot$CH$_3$). All these but the last two are products of the breakdown of organic material, usually bacteria or fungi, although plants can also release these volatiles. DMS is produced by marine organisms like a breakdown product of the DMSP (dimethylsulfoniopropionate) generated by marine plankton perhaps for  stress resistance (\cite{seageretal2012} and reference therein). Much of the DMS generated is consumed by other organisms but part is released in the atmosphere. This gas, under the right conditions of excess production or favorable ultraviolet (UV) flux conditions, could accumulate to potentially detectable levels.  Moreover, it is worth to mention that dimethyl sulfide is the largest biological source of atmospheric sulfur.

At the end of the description of this kind of biosignature it is necessary to rise a caveat. Unlike the products of primary metabolism, we cannot predict the circumstances in which secondary metabolism by--products might be produced on other worlds. The strength of this kind of biosignature is in the lack, almost complete, of false positive which plagues the biosignatures of type I.

\subsection{Bioindicators}
\label{sect:bioindicators}
Other gases that are or could be indicative of biological processes, but can also be produced abiotically, or biosignature gases that can be transformed into other chemical species abiotically are considered bioindicators. 

In the list of biosignatures water (H$_2$O) doesn't appear also if it is a by--product of many reactions generating biosignatures.  This is because water, as well as with CO$_2$, could not be considered as sign of life itself, but it is raw material for life and an important molecule for planetary habitability as a greenhouse gases. For example, on an Earth-like planet where the carbonate-silicate cycle is at work, the level of CO$_2$ in the atmosphere depends on the orbital distance; CO$_2$ is a trace gas when the planet is close to the inner edge of the HZ but a major compound in the outer part of the HZ (\cite{koppaparuetal2013}). Other kinds of gases could be considered in this ensemble, for example both SO$_2$ and H$_2$S, a gas mixture produced by volcanism and out of thermodynamic equilibrium at terrestrial surface conditions. The reactions between the two will form water and elemental sulfur. Detecting both H$_2$S and SO$_2$ in an exoplanet atmosphere could therefore be either a sign of life or just a sign of volcanism \cite{lammeretal2009}. Other bioindicators reported by several authors include ethane (a hydrocarbon compound) from biogenic sulfur gases (\cite{domagalgoldmanetal2011}) and hazes generated from CH$_4$ (\cite{haqq--misraetal2008}).

In case of transformation of a biosignature by abiotic process, the resulting product might also not be naturally occurring in a planet's atmosphere and therefore also a sign of life. One of the most important examples is due to the photochemistry of O$_2$ that produces ozone (O$_3$). Once that oxygen reaches the high part of the atmosphere, between 14 and 30 km in Earth's atmosphere, it could be modified by the UV radiation coming from the host star.
The photochemical production and destruction of ozone are then only governed by the Chapman cycle (\cite{chapman1930}):

$$O_2 + h\nu_1 \rightarrow O + O$$
$$O + O_2 + M \rightarrow O_3 +M$$
$$O_3 + h\nu_2 \rightarrow O_2 +O$$
$$O + O_3 \rightarrow 2 O_2$$
$$ O + O + M \rightarrow O_2 + M$$

\noindent
In the set of Chapman's reaction it is important the UV radiation, in particular $h\nu_1$ is the energy of photons in the range between $0.1 - 0.2$\ $\mu$m, while $h\nu_2$  are photons in the range between $0.2 - 0.3$\ $\mu$m. M is any molecule, mostly O$_2$ and N$_2$ on Earth. This reaction is not very efficient as it requires at the same time a high enough pressure (because it is a 3 body reaction), and oxygen atoms that are produced at lower pressures where photolysis of O$_2$ by UV can occur (\cite{kalteneggerandselsis2007}). Ozone can be efficiently destroyed by a large number of reactions dominated, in the Earth's atmosphere, by catalytic cycles involving species such as hydrogenous compounds (H , OH, HO), nitrogen oxides (NO$_X$) and chlorine compounds (ClO$_X$). These species have various origins and their amount depends on the nature and the intensity of the bio--productivity, the thermal profile of the atmosphere, human pollution, and many other parameters. Without these compounds, an atmosphere made of N$_2$ and O$_2$ only would contain 10 times more O$_3$. The column density of O$_3$ in the atmosphere depends weakly on the abundance of O$_2$, the mean opacity of the 9.6 $\mu$m band remaining $> 1$ for O$_2$ abundance as low as 10$^{-3}$ present atmospheric level (\cite{legeretal1993}, \cite{seguraetal2003}).  

The bottom line is that ozone is a tracer of the O$_2$. Leg\'er et al (\cite{legeretal1993}) modelled the production of ozone in atmosphere with the presence of O$_2$ and studied  the variation of O$_3$ column density at the variation of O$_2$ amount in atmosphere. O$_3$ is a non--linear tracer of O$_2$ because its spectral features become rapidly saturated. The depth of the saturated O$_3$ band is determined by the temperature difference between the surface clouds continuum and the ozone layer. In any case the visible and IR features of O$_3$ are easier to be observed that O$_2$ features.

\subsection{Surface and Industrial Biosignatures}
\begin{figure}
\begin{center}
 \includegraphics[width=.7\textwidth]{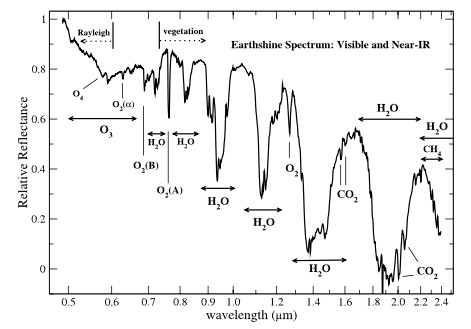}
\caption{ Observed reflectivity spectrum of Earth as determined from Earthshine in the range spanning from Visible to NIR. The reflectivity of vegetation containing chlorophyll, on land, is dominated by a sharp rise in reflectivity for wavelengths longer than about 0.72 $\mu$m, plus some smaller bumps at shorter and longer wavelengths. (\cite{woolfetal2002}; \cite{turnbulletal2006}). In the spectrum, the rise of the flux due to plants reflection is only about 6\% of the nearby continuum. This is because at the time of observation only about 17\% of the projected area was land (for details on the observing method see \cite{woolfetal2002}). In the spectrum are also indicated most of the molecules that are addressed in the text.}
 \label{fig:earthshine}
\end{center}
\end{figure}

In order to complete the description of possible biomarkers, we can consider other possible sign of life not included in the previous classification of biosignatures. They can be called surface and industrial biosignatures. The former class derives by the presence on the planetary surface of vegetables and their pigments that reflect the incident light coming from the host star with a peculiar reflectance spectrum that shows a rise at about 700 nm. This shoulder is called ''red--edge'' and it is distinctive of vegetation. Physical explanations of land plant spectral signatures are fairly well understood in some aspects, whereas there is less of such information on other photosynthesizers. Technically, the red--edge is a spectral reflectance feature (e.g. see Figure\ \ref{fig:earthshine}) characterized by darkness in the red portion of the visible spectrum, due to absorption by chlorophyll, strongly contrasting with high reflectance in the NIR, due to light scattering from refraction along interfaces between leaf cells and air spaces inside the leaf (\cite{tinettietal2006}).

In their paper, Lin et al. (\cite{linetal2014}), by re--elaborating the idea of Owen (\cite{owen1980}), pointed out that in addition to these generic indicators, anthropogenic pollution could be used as a novel biosignature for intelligent life. In particular they focused on chlorofluorocarbons (CFCs): tetrafluoromethane (CF$_4$) and trichlorofluoromethane (CCl$_3$F), which are the easiest to detect of the molecules produced by anthropogenic activity. The main spectral signatures of these molecules are in the range between 7760 $< \lambda <$ 7840 nm in the case of CF$_4$ and 11600 nm $< \lambda < $12000 nm for CCl$_3$F. Their abundances are too low to be spectroscopically observed at low resolution (\cite{seguraetal2005}). In any case Lin et al. studied the case of an Earth in the HZ of a white dwarf (high transit probability) like Agol and try to foreseen the possibility of CFCs detection with JWST. They estimated that $\sim 1.2$ days ($\sim 1.7$ days) of total integration time will be sufficient to detect or constrain the concentration of CCl$_3$F (CF$_4$) to $\sim10$ times current terrestrial level.

\section{False Positives}
Most of the features described in the previous sections as biosignatures actually are not unique by--product gases due to the presence of life. There are a lot of atmospheric and geophysical processes that are able to produce the same kind of molecules in detectable quantities. The most prone to false positives are type I biosignatures. In this case, in fact, geology uses the same redox gradient in order to produce the same molecules produced by life. If the planet is geologically active, hot spots, volcanism, fumaroles and hot springs are the main actors that are able to produce CO, CO$_2$, CH$_4$, H$_2$O, N$_2$ H$_2$S. The last one is produced by volcanos in large amount. Moreover there are gases, like N$_2$ and H$_2$O that are by--products of life and that are present in considerable amount in the Earth's atmosphere.  For example on Earth N$_2$ makes up 80\% of our atmosphere and it is essentially a  product of primordial outgassing derived from geological sources during the planet formation. Water was also present on Earth since the beginning. 

The most important false positive is the abiotic production of oxygen. The main abiotic routes for oxygen are both coming from photochemical reactions due to the photodissociation of CO$_2$ and H$_2$O. The CO$_2$ photolysis is due to UV radiation coming from the host star with a wavelength of about 140 nm:

$$ 2(CO_2 + h \nu) \rightarrow 2(CO +O)$$

it is followed by a recombination of oxygen:

$$ O+O+M \rightarrow O_2+M$$

with the net result of the loss of two CO$_2$ molecules and the production of two CO molecules and one O$_2$. To reach detectable levels of O$_2$ (in the reflected spectrum), the photolysis of CO$_2$ has to occur in the absence of outgassing of reduced species and in the absence of liquid water because of the wet deposition of oxydized species. Normally, the detection of the water vapour bands simultaneously with the O$_2$ band can rule out this abiotic mechanism (\cite{seguraetal2007}), though one should be careful, as the vapor pressure of H$_2$O over a high-albedo icy surface might be high enough to produce detectable H$_2$O bands (\cite{kalteneggeretal2010b}). In the atmospheres of Venus and Mars, the photolysis of CO$_2$ is a source of atomic oxygen.

\begin{figure}
\begin{center}
\includegraphics[width=.7\textwidth]{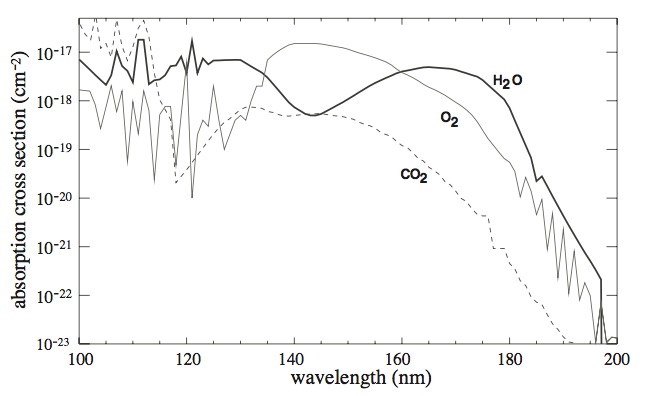}
\caption{Cross section for photodissociation of H$_2$O, CO$_2$, O$_2$. Taken by Selsis et al. (\cite{selsisetal2002}).}
\label{fig:crosssection}
\end{center}
\end{figure}

The photodissociation of water, occurs instead when the planet is under a runaway greenhouse effect due to a strong warming of the atmosphere. Liquid water on the surface of the planet is vapourized adding greenhouse gas to the already present gas in the atmosphere. The atmosphere becomes warm and moist and the temperature inversion layer reaches a higher altitude in the atmosphere, causing water vapour to fall prey of the UV radiation. Around 140 nm, H$_2$O absorbs UV photons in the same wavelength range as CO$_2$ (see Figure\ \ref{fig:crosssection}), and it is photo--dissociated by the following reactions:

$$4(H_2O+h\nu)\ \rightarrow \ 4(OH+H)$$

$$2(OH+OH)\ \rightarrow \ 2 (H_2O+O)$$

the oxygen reacts with itself in order to reproduce one molecular oxygen

$$O + O\ \rightarrow \ O_2$$

while the light hydrogen escapes form the gravitational pull of the planet. The net result of this photochemical process is the destruction of four water molecules with the production of one molecule of oxygen and the loss of four hydrogen atoms to space. This situation could lead to detectable O$_2$ levels (\cite{kasting1988}).

In any case, the two processes are strongly coupled: in fact the efficiency of O$_2$ production from H$_2$O photodissociation decreases with increasing CO$_2$ abundances, as CO$_2$ absorbs UV photons in the same wavelength range as H$_2$O. 
Moreover the photochemical production of oxygen is quite self-regulating, because O$_2$ could be also dissociated by the same photon that splits H$_2$O and CO$_2$ (\cite{selsisetal2002}). 

Thus, abiotic O$_2$ production will be more efficient either in a CO$_2$ dominated atmosphere with very little or no water or in a humid atmosphere poor in CO$_2$ and the loss of hydrogen from the atmosphere into space can result in huge leftover of oxygen. As matter of fact, Venus shows us that this huge quantity of oxygen due to the photolysis of water experienced by the planet in the past has a limited lifetime in the atmosphere due, for example, to the oxidation of crust and to the oxygen loss into space. The loss of hydrogen and the photochemical induced production of oxygen is driven by the distance from the host star and from the gravitational pull of the planet itself (\cite{kastingetal1993}, \cite{selsisetal2007}, \cite{koppaparuetal2013}). Less massive planets close to their star experiencing runaway greenhouse effect can lose water easier than heavier and farther planets. To a less extent, this process could work also in Earth--like planets warning us that life is not the only process able to enrich an atmosphere with these compounds.

\section{Methods and Instruments}
The search for life in other site than Earth could be done with two different methods: via site exploration and via remote sensing. The former is the method used for the planets, satellites and small bodies of the Solar System. A lot of NASA, ESA and Russian Space Agency missions had and will have as goal the exploration of Venus, Mars, Titan and other Solar System bodies. This method,  quite difficult and expensive also for the closest objects of the Solar System, could not be used for the discovered extrasolar planets, also in the case of Proxima Cen b (\cite{angladaescudeetal2016}) that is ''only'' at 1.295 pc from the Sun thus the remote sensing seems, so far, the only possible method. As told in Section\ \ref{sec:intro}, the approach to remote detection of signs of life on another planet was set out in \cite{lederberg1965} and \cite{lovelock1965} over one--half a century ago. Since the first discovered planet, there was a lot of effort in building instruments more and more sensitive, able to record the very small reflex motion induced on the host star by the presence of planetary  bodies (e.g. \cite{mayoretal2003}, \cite{cosentinoetal2012}). In the meantime the same efforts were done in order to improve the possibility to directly observe the planet. In both cases the improvements brought to began to probe, for few particular cases, the atmosphere of planets. 

Direct imaging (\cite{claudi2016}, \cite{daviesandkasper2012}) is specialized, for the time being, on young giant planets orbiting far away from the host star. This method is limited by the available angular resolution (due to the telescope diameter and adaptive optics) and by the necessity to eliminate the glare of the star using coronagraphs. Coronagraphs limit physically the possibility to observe regions very close to the star and, also if the new generation of these optical devices is able to work with very small IWAs (Inner Working Angle \footnote{IWA is universally defined as the 50\% off--axis throughput point of a coronagraphic system, expressed usually in $\lambda/D$ (resolution element)}), the HZ of stars distant more than 10 pc are out of reach. In fact, the smallest IWA possible is  about $2 \lambda/D$ (\cite{mawetetal2012}) that means 0.8 au at 10 pc. Note that the inner border of HZ of the solar system is at about 0.9 au (\cite{kastingetal1993}, \cite{koppaparuetal2013}). Other important difficulties affect the possibility of getting directly the spectrum of a planet: first of all there is the capacity to observe the smaller planets at the high contrast of flux of $10^{-10}$, required for an Earth--like planet (see e. g.  \cite{chauvinetal2005}, \cite{lafreniereetal2007}, \cite{billeretal2013}). 

SPHERE@VLT(\cite{beuzitetal2008}) and GPI@GEMINI (\cite{macintoshetal2014}), the ultimate high contrast imagers, that have scores on detection of young giant planets (e.g. \cite{maireetal2016}, \cite{zurloetal2016}) in the outer regions of host stars are the forerunners for similar instruments to be mounted on the new generation of extremely large telescope in construction: E--ELT (European--Extremely Large Telescope), Thirty Meter Telescope (TMT) and Giant Magellanic Telescope (GMT). These telescopes and their instrumentation will be able, in the next future, to observe and perhaps obtaine the spectra of the nearer exoplanets orbiting in the HZ of their stars.

Now, the more productive technique for atmosphere observation takes advantage of the combined light of transiting planets and their stars. The technique could be split in the following: transit transmission spectra (\cite{charbonneausetal2002}, \cite{seagerandsasselov2000}) and secondary eclipse spectra in thermal emission (\cite{charbonneausetal2002} , \cite{demingetal2005}) and reflected light (\cite{demoryetal2011}, \cite{evansetal2013}). Transmission spectroscopy, possible only when the planet transits its host star along the line of sight, allows to infer the main opacity sources present in the high atmosphere of the planet (\cite{brown2001}, \cite{tinettietal2007}). Complementary, emission spectroscopy (\cite{charbonneauetal2005}), observing the day hemisphere of the planet and exploiting its occultation during the secondary transit, gives evidence on the thermal structure of the planetary atmosphere and the emission/reflection properties of the planetary surface. Ground-based atmospheric characterisation of exoplanets advanced through the use of high-resolution spectrographs like High Dispersion Spectrograph (R = 45,000) at the Subaru 8-m telescope, CRIRES (Cryogenic High-Resolution Infrared Echelle Spectrograph) and its new refurbishment CRIRES + at VLT. Also high resolution spectrograph at smaller diameter telescope but with larger spectral range like GIARPS (GIAno and haRPS) at Telescopio Nazionale Galileo could be proficient in this business (\cite{claudietal2016}, \cite{carleoetal2016}). With these kinds of spectrographs it is also possible to use an alternative technique to characterise the planets' atmosphere. In fact, for those planets with short orbital periods and resulting high orbital velocities, a high spectral dispersion cross-correlation technique could be (and actually has been) used to measure atmospheric spectral features, taking advantage of the planet's orbital motion (with the subsequent Doppler shift in the planetary spectrum compared to the stellar spectrum) and a known template of high spectral resolution molecular lines (\cite{snellenetal2010}). So far, these techniques were applied with success to hot Jupiter and other giant planets, but Bean et al (\cite{beanetal2010}) observed spectroscopically the warm super Earth orbiting the M star GJ 1214b finding a flat transmission spectrum, due to the strong irradiation of the host star on the surface or to the presence of optically thick clouds, or both.

Some results in the past were obtained with space--based observations of hot Jupiters using HST and Spitzer (e.g. \cite{seageranddeming2010}) and, on the way,  there are several space missions of ESA and NASA that are going to be launched in space in order to find and characterize extrasolar planets. Sorting by starting date of the mission, CHEOPS (CHaraterising ExOPlanet Satellite \cite{broegetal2013}) will measure precise radii and bulk densities of known exoplanets by transit photometry and select the best-suited targets for atmospheric characterisation with future spectrographs from space or on large ground-based telescopes. The launch of this mission is expected for the 2018. Other missions, with demographic vocation, are going to be launched following the pathway initiated by CoRot (\cite{moutouetal2013}) and uppermost {\it Kepler} (\cite{boruckietal2009}). TESS (Transiting Exoplanet Survey Satellite) developed by NASA, (\cite{rickeretal2009}) and PLATO (Planetary Transits and Oscillations of Stars) by ESA (\cite{raueretal2016}) will monitor photometrically, bright stars and thus find many new transiting planets around bright stars adding a lot of target of which will be possible to characterise the atmosphere. 

NASA's and ESA's James Webb Space Telescope (JWST; launch expected in 2018) will enjoy an unprecedented thermal infrared sensitivity and provide powerful capabilities for direct imaging, including coronagraphy (\cite{gardneretal2006}).  It will mount four instruments: a short-wavelength imager NIRCam, NIRISS, a complementary imager that utilise sparse Aperture mask (SAM) in the wavelength range between $1-2.3\ \mu$m, MIRI, the spectrograph  in the $5-28\ \mu$m wavelength range, and finally NIRSPEC ($1-5\ \mu$m) will be equipped with an integral field spectrograph. Its four instruments will, in addition to direct imaging of planets, attempt transit observations at low--to medium--resolution ($100 < $R$ < 1500$) in the near- and mid-infrared domain for atmospheric characterisation. The synergy between the discovery possibilities of TESS and the capability of JWST will allow to characterize several super Earths among with some in the HZ and life detection is a possibility if life turns out to be ubiquitous on exoplanets (\cite{seager2014}).

Finally, it is worth to mention the proposed mission ESA--M5 ARIEL (Atmospheric Remote-Sensing Infrared Exoplanet Large-survey) already in its infancy, that will conduct a large, unbiased survey of exoplanets in order to begin to explore the nature of exoplanet atmospheres and interiors and, through this, the key factors affecting the formation and evolution of planetary systems (\cite{tinettietal2016}). ARIEL, that will be fully dedicated to this aim, will carry a single, passively-cooled, highly capable and stable spectrometer covering $1.95 - 7.80\  \mu$m with a resolving power of about 200 mounted on a single optical bench with the telescope and a Fine Guidance Sensor (FGS) that provides closed-loop feedback to the high stability pointing of the spacecraft. ARIEL will observe a large number ($\sim 500$) of warm and hot transiting gas giants, Neptunes and super--Earths around a range of host star types using transit spectroscopy in the $\sim2-8\ \mu$m spectral range and broad-band photometry in the optical.

\section{Summary and Conclusions}
In the previous sections tentative answers  to the three questions {\it where, what} and {\it how} to search for life out of there are given. First of all, rocky planets, named super Earths, seems to be the really putative locii where to search for alien life. What we can search for in the atmosphere of super Earths is also addressed in the previous sections. While the former is today a straightforward answer, the latter is complicated and not so clear. In fact, we have considered different classes of spectral features coming from life metabolic reactions working on Earth. We discussed also the possible false positives that can be generated by geo-physical and geo--chemical reactions. In Table\ \ref{tab:biosigsummary}  the different types of biosignatures are summarised together with the pro and con's for their detection. 

\begin{table}[h]
\begin{center}
\begin{tabular}{ccll}
\hline
Biosignatures            & Example                          & Pros                                   & Cons \\
\hline
\hline
Inorganic Gases        & O$_2$; O$_3$; CH$_4$ & Detectable                         & Also abiotic \\
(Type I, Type II)         &                                          &                                           & process       \\
\hline
Organic Gases          & DMS; isoprene etc.          & Only bio by--product          & Small Concentration\\
(Type III)                    &                                         & Many different Compounds&                     \\
\hline
Photopigments          & Chlorophyll                      &       Unique                          & Complex in detecting  \\
 (surface reflectance)&                                         &                                             &and for evolution      \\
\hline
Industrial                   & CF$_4$; CCl$_3$F         & complex evolved             & Small Concentration \\
                                  &                                         & and polluting life             &                                   \\
\hline
\hline
\end{tabular}
\caption{Summary of biosignature classes with pros and cons}
\label{tab:biosigsummary}
\end{center}
\end{table}%

\noindent
On Earth, the only sure biosignatures netted off false positives are O$_2$ and N$_2$O.
The other primary metabolism gas by--products discussed are either short-lived in the atmosphere or also produced in dominant amounts by geophysical processes.  For the present day, methane is not readily identified with low resolution due to the low atmospheric abundance. In any case it is worth to mention that at higher abundance, as it was in the primeval Earth (\cite{kastingandcatling2003}, the methane IR feature at $7.66\ \mu$m will be easily detectable (\cite{lammeretal2007}). A very promising biosignature could be the DMS, the secondary  metabolic by--product discussed in the description of type III biosignature class, but in that case, on Earth, we have to compel with small atmospheric concentration. On an extrasolar planet could be also more difficult that this kind of high specialised by--products could be synthesised by organisms that evolved in a different way and under different environment condition of those on Earth.

So far, O$_2$ has been assumed as the best case for a biosignature gas in the search for life beyond our solar system, and the presence into the atmosphere of its photosinthetized product O$_3$ is considered as the evidence of the presence of life forms producing oxygen. In any case we have to pay attention to the retrieved spectra in order to unambiguously identify O$_2$ and other atmospheric gases, which would set the environmental context in which we are confident that the O$_2$ is not being geochemically or photochemically generated (\cite{domagalgoldmanetal2014}). Actually Ozone is a not a linear indicator of the presence of biotic oxygen. In fact a saturated ozone band appears already at very low levels of O$_2$ ($10^{-4}$ ppm), while the oxygen line remains unsaturated at values below the present atmospheric level (\cite{seguraetal2003}). In addition, the stratospheric warming decreases with the abundance of ozone, which makes the O$_3$ band deeper for an ozone layer less dense than that in the present atmosphere. The depth of the saturated O$_3$ band is determined by the temperature difference between the surface-cloud continuum and the ozone layer. The O$_2$ has a strong visible signature at $0.76\ \mu$m (the Frauenhofer A-band) observable with low/medium resolution and also a faint one at $0.69\ \mu$, visible with high resolution (\cite{kalteneggeretal2010b}, \cite{tinettietal2013}). 

N$_2$O is detectable at 7.8, 8.5 and 17.0 $\mu$m with high resolution (\cite{kalteneggeretal2010b}, \cite{tinettietal2013}). It would be hard to detect in the Earth's atmosphere with low resolution, as its abundance is low at the surface (0.3\ ppmv) and falls off rapidly in the stratosphere.  Detection of water vapour in a planetary atmosphere is the first clue indicating that a planet may be habitable. 

Water absorption can be seen in several bands in the visible and infrared. On the visible-NIR ($0.5-1\ \mu$m) it is possible to detect water absorption at 0.7, 0.8 and $0.9\ \mu$m. Other possibilities are the bands at 1.1, 1.7 and $1.9\ \mu$m. As for the mid-IR, water can be detected between 5 and $8\ \mu$m and from around 17 out to $50\ \mu$m.  With extremely high resolution, the absence of water can also be deduced from the presence of highly soluble compounds like SO$_2$ and H$_2$SO$_4$. Venus' spectrum, for example, has some weak absorption bands from water but its bulk atmospheric chemistry can only be explained if the planet is dry.

To search for life on other planets is important to characterise the atmosphere of the planet in order to understand how it works from both chemical and physical point of view. In order to do that we have several useful tools like e.g. the definition of HZ or the understanding of how life as we know it works on a planet and what kind of by--product it synthesise.  In any case, this is only an half of the story. In fact if we are going to search for life without any constraints on what kind of life we are looking for, we need to tackle this argument with an open mind and, for example, regardless of the possibility that the planet is in the HZ or not, searching for no--equilibrium quantities. Once stated this, we have to go deep and investigate the reasons of that.  

How many years will be necessary to obtain some hints on it? It's difficult to answer to this question. At the moment we have several brand new instruments both ground and space based, that are devoted to the aim to find and characterise extrasolar planets. In the next future, huge ground telescopes, a dedicated space mission and one space telescope (JWST) will be able to begin to address such an intriguing astrophysical problem.

\section{Acknowledgesments}
\noindent
Author acknowledges support from INAF trough the ''Progetti Premiali'' funding scheme (WOW project) of the Italian Ministry of Education, University, and Research. The author thanks a lot Eleonora Alei, Serena Benatti and Ilaria Carleo for the useful discussions and suggestions and lastly, P.A. Mason for referring this review and for the very useful help.

\end{document}